\documentstyle[11pt]{article}
\newcommand{\blankline}{\vskip .3cm}
\newcommand{\f}{\begin{equation}}
\newcommand{\ff}{\end{equation}}

\begin{document}
\rightline{\Large CGPG-93/9-4}
\vfill
\centerline{\LARGE Fermions and topology}

\blankline
\rm
\vskip1cm
\centerline{Lee Smolin${}^*$}
\blankline
 \centerline{\it   Center for Gravitational Physics and Geometry}
\centerline{\it Department of Physics}
\centerline{\it The Pennsylvania State University}
 \centerline {\it University Park, Pennsylvania, 16802-6360   U.S.A.}
 \vfill
\centerline{March 24, 1994}
\vfill
\centerline{\it Dedicated to John Archibald Wheeler}
\centerline{ABSTRACT}
\blankline
\noindent
The canonical theory of quantum gravity in the loop representation
can be extended to incorporate topology change, in the simple
case that this
refers to the creation or annihilation
of "minimalist wormholes" in which two
points of the spatial manifold are identified.   Furthermore, if
the states of the wormholes threaded by loop states
are taken to be antisymmetrized
under the permutation of wormhole mouths, as required by the
relation between spin and statistics, then the quantum theory
of pure general relativity, without matter but with minimalist
wormholes, is shown to be equivalent to the quantum
theory of general relativity coupled to a single Weyl fermion field,
at both the kinematical and diffeomorphism invariant levels.
The correspondence is also shown to extend to  the
action of the dynamics generated by the Hamiltonian constraint,
on a large subspace of the physical state space, and is thus
conjectured to be completely general.
\vfill
${}^*$ smolin@phys.psu.edu
\eject

\section{Introduction}

The idea that matter is made up out of the geometry of space is older
than relativity theory, as it was discussed in the 19th Century by
Clifford\cite{history}.
In the context of relativity theory a new possible arises,
which is that matter can be constructed out of the toplogy of space.
This idea has been championed by John Wheeler, who
with Charles Misner many years ago proposed that
wormholes at the Planck scale might behave as
charged particles\cite{JAW-wormholes} and
by Friedman and Sorkin, who showed that when the three manifold
has non-trivial topology the quantum states of the gravitational field
can have half integral angular momentum\cite{johnrafael}.

In this note I would like to show that the idea of Wheeler can be
realized in a particularly straightforward way in the loop
representation formulation of non-perturbative
quantum gravity
\cite{review,review-ls,abhay-leshouches,abhay,carlolee,gambini}.
What I will show is that a kind of minimal wormholes can be
precisely identified as fermions in the sense that one can
construct an
isomorpism between the states and operators of the theory of
general relativity coupled to fermions and a corresponding set of
states of pure general relativity, with no matter, but with wormholes
permitted.  This correspondence can be established both at the
kinematical and the dynamical levels.

The basic idea of the construction I will describe here is to make the
description of the wormholes as simple as possible.  The simplest
possible wormhole can be constructed just by picking two points of
the three manifold, $\Sigma$,  which we may call $x$ and $y$, and
identifying them.  Such
a wormhole may be called a minimalist wormhole.
 Of course, we loose the Hausdorf property, but in
such a mild way that it is still possible to define the structures that
are needed to construct general relativity at both the classical and
the quantum level.  Indeed, spaces with such
identifications are called conical spaces by the
mathematicians, and have been well studied.
Whether this construction is seen as
fundamental, or as a simplification that can allow the details of the
construction of the wormhole to be neglected and that might be
eventually supplemented by details about the short distance
properties of the wormhole, is left to the taste of the reader.

In the loop representation, states with fermions are represented as
open loops, which stand for a holonomy element tied up with a
fermion at each end\cite{loop-fermions,carlohugo}.
A closed loop that, however, goes through two
identified point looks like an open loop; this is the basic idea of the
identification of states with fermions with states of pure gravity on a
manifold with identified points.

The first task facing such a construction is the incorporation of
topology change in a canonical quantum theory.  In the case of
minimalist wormholes this is particularly straightforward.
For clarity, I will develop first the quantum theory of minimilist
wormholes, without mentioning fermions.    Then I will show
that the operators that describe creation, annihilation and motion
of wormholes have precisely the same algebra as those operators
that are used to describe
fermions in quantum gravity.   Having thus established
the correspondence at the kinematical level, I will
then show that it can be extended to the
level of  diffeomorphism invariant invariant states.
Following this, I show that for certain classes of states the
correspondence extends to the level of the  action of the
Hamiltonian constraint that generates the
dynamics.  There may in fact be a  complete  equivalence
between fermions and these minimalist wormholes, but to show
this for general states remains a problem for future work.

\section{Quantum kinematics of topology change}

In order to develop the quantum theory of minimalist wormholes
we need operators that take us between states defined on spaces
with different topologies.
This is the main task of this section\footnote{I will work here in
3 dimensions, but the construction works in other
dimensions.}.

To begin with we need some notation and
mathematical preliminaries.  Beginning with a proper
differentiable manifold $\Sigma$, we may consider the
conical manifold defined by identifying two points $x$ and $y$
of $\Sigma$.  I will call this $\Sigma_{(x,y)}$.  It is defined by
restricting the commutative algebra of $C^\infty$ of functions
$f$ on $\Sigma$ to the subalgebra in which $f(x)=f(y)$.  The
vector fields on $\Sigma_{(x,y)}$ may be defined following the
usual route from the derivations on the function algebra, from
which the various other structures are defined.

Similarly, we
may define the manifold
$\Sigma_{(x_1,y_1)...(x_n ,y_n)}$
with $n$ pairs of points, $x_i, y_i $ identified
by requiring that for all $  i, f(x_i) = f(y_i)$.

In any quantum theory of interest we will have a space of states,
${\cal H}_{(x_1,y_1)...(x_n ,y_n)}$ defined on each
$\Sigma_{(x_1,y_1)...(x_n ,y_n)}$.  In some quantum field theories,
such as those defined by Fock constructions, there may be subtelties
concerning defining the states on the minimilist wormhole manifolds.
But we will here be concerned with state spaces appropriate to
the non-pertubative quantization of diffeomorphism invariant
theories, principly the constructions based on using the discrete
norm on the space of loops\cite{rayner,review-ls},
as formalized by Ashtekar
and Isham\cite{abhaychris}.  These state spaces have bases
that, in the connection
representation, may be represented as countable
sums of products of holonomies, or in the loop representation
have support on countable sets of loops. We may note that the
recent work of Ashtekar and Lewandowski\cite{abhayjerzy} and
Baez\cite{baez} establishes the existence of diffeomorphism invariant
cylindrical measures on these spaces, elevating the heuristic
constructions of previous work to the level of rigor of
perturbative quantum field theory in Fock representations.
For these kinds of representations there
are, as we shall see,
 no difficulties associated with the non-Hausdorff behavior
at minimalist wormholes.

Thus, for each $\Sigma_{(x_1,y_1)...(x_n ,y_n)}$ the state space
${\cal H}_{(x_1,y_1)...(x_n ,y_n)}$ will be spanned by the
states $ <\alpha ; (x_1,y_1)...(x_n ,y_n)|$, which satisfy all the
identities of states in the appropriate loop
representation
\cite{review,review-ls,abhay-leshouches,abhay,carlolee,gambini}
\footnote{From now on I will use notation
appropriate to the loop representation, however all the
definitions and identities may be taken as referring to the
corresponding connection representation, with the connection
states $\Psi_{\alpha } [A]$ put on the right, instead of their adjoint
loop states $<\alpha ;...|$ on the left.}.  In
particular, the $\alpha$ must contain only closed loops, although
the cases in which a loop is closed by virtue of its passage through
a minimalist wormhole defined by two identified points are
included.

The total Hilbert space of interest may be then written as
the direct sum of all these spaces, thus
\f
{\cal H}_{total} \equiv \sum_{n=0}^\infty
\bigoplus_{x_1}...\bigoplus_{x_2n}
{\cal H}_{(x_1,x_2)...(x_{2n-1}x_{2n})}.
\ff
The continuous products here are defined using the discrete norm.
We may note
that in this formalism the order that the wormholes are
created doesn't matter, a point may be at the mouth of
more than one wormhole and the case of a minimalist wormhole
connecting a point $x$ to itself is allowed.

We now need operators that connect states in the different
state spaces ${\cal H}_{(x_1,x_2)...(x_{2n-1}x_{2n})}$.  If
$< \alpha ; (x_1,x_2)...(x_{2n-1}x_{2n})|$ denotes the bra
associated with the (possibly multi-) loop $\alpha$ in
${\cal H}_{(x_1,x_2)...(x_{2n-1}x_{2n})}$ we may define
the operator $\hat A(y,z)$ by
\f
< \alpha ; (x_1,x_2)...(x_{2n-1}x_{2n})| \hat A (y,z) =
< \alpha ; (x_1,x_2)...(x_{2n-1}x_{2n})(y,z)|
\ff
We may note that there is no difficulty if the loop $\alpha$
include the points $y$ or $z$, the state on the right hand side
then corresponds to loops that pass the identified point without
making use of the option given by the
identification of continuing "on the other side" of the wormhole.
We may also note that as the classical canonical theory cannot
describe
topology change these operators do not correspond to any
classical observables.  We will, however, see below that they can
be interpreted to describe the creation of a pair of fermions.

The minimalist
wormholes are unoriented, so that $\hat A(y,z)=\hat A(z,y)$.
It makes no difference in which order they are created, so
distinct wormhole creation operators commute.  However,
it makes no sense to identify two points twice, therefor we
impose,
\f
\hat A(y,z)^2 =0
\ff

We may also define two other sets of useful operators.
$\hat B(z \rightarrow w)$ is the operator that moves a mouth
of a wormhole from $z$ to $w$.   That is, it is a map from
${\cal H}_{(x,z)...}$ to ${\cal H}_{(x,w)...}$ defined by
\f
<\alpha ; (x_1,x_2)...| \hat B( z \rightarrow w) \equiv
\delta^3 (z, x_1 ) <\alpha ; (w,x_2)...| +
\delta^3 (z, x_2 ) <\alpha ; (w,x_1)...|
\ff
This implies that,
\f
\left [ \hat B (z \rightarrow w ), \hat A(x,y) \right ]
= \delta^3 (z,x) \hat A(w,y) +
\delta^3 (z,y) \hat A(w,x) .
\ff
We may also define a projection operator onto the space
${\cal H}_{(x_1,x_2)...(x_{2n-1}x_{2n})}$, which will be
denoted $\hat{P}[(x_1,x_2)...(x_{2n-1}x_{2n})]$.  We may note
that it satisfies
\f
\hat A(x_1, x_2 )\hat{P}[(x_1,x_2)...(x_{2n-1}x_{2n})] = 0
\ff
and
\f
\hat{ P}[(x_1,x_2)...(x_{2n-1}x_{2n})] \hat A (x_1 , x_2 ) =
\hat A (x_1 , x_2 )
\ff

Because all observables must be constructed from the field
operators
of the quantum theory, a wormhole through which no loop
passes will be unobservable.  Thus, we will require that the
states be restricted to superpositions of those in which there
are no such "naked wormholes".
To accomplish this we couple
the action of the wormhole creation operators $\hat A(x,y)$ with
operators that create loops that go through the minimalist wormhole
gotten by identifying $(x,y)$.  Thus, for every open segment
$\beta$ in $\Sigma$ we may define,
\f
\hat{\bf T} [\beta ]\equiv \hat T [\beta ] \circ \hat
A[\beta (0) , \beta (1)]
\ff
We will then restrict ourselves to states  reached from
states in the state space ${\cal H}$ defined on $\Sigma$
without identifications by the action of these
operators.  These spaces will be denoted by a prime as
in
${\cal H}_{(x_1,x_2)...(x_{2n-1}x_{2n})}^\prime $.

It is natural to have a dressed version of the operator
$\hat B [z \rightarrow w ]$ that moves a wormhole
mouth from one point to another and adds an appropriate
loop segment connecting the old point to the new one.
This may be written as, $\hat B [\beta]$, where $\beta$
is an oriented loop segment in $\Sigma$, and defined by
\f
\hat B[\beta ] \hat {\bf T} [\gamma ] \equiv
\delta (\beta (1), \gamma (0)) \hat {\bf T} [\beta \circ \gamma ]
+
\delta (\beta (1), \gamma (1))
\hat {\bf T} [\beta \circ \gamma^{-1} ].
\ff

We may note that it is difficult to write these directly in
the loop representation, as there are no operators
that remove loops (at least that correspond to known
classical observables).  But a classical observable
corresponding to $\hat{B} [\beta ]$ can be found in
the theory in which wormhole mouths
are identified as fermions, as we will see below.

To complete the definition of the loop or connection
representation we need to include operators that are linear in
momentum variables. This allows us to construct a closed
algebra of elementary obserables whose representations
define the quantum theory.   There are, first of all, the standard
operators $\hat{T}^a[\gamma ](s)$ (which may be
regularized in terms of the strip operators).  In addition,
there are "$T^1$" operators analogous to
$\hat{\bf T}[\beta ]$ and $\hat B [\beta ]$ assoicated with open
curves $\beta $ in $\Sigma$ that may be defined by
their actions on ordinary $T[\alpha ]$'s by
\f
\left [ \hat {\bf T}^a [\beta ](s) , \hat T [ \alpha ]  \right ]
\equiv
\int dt \delta^3 (\beta (s), \alpha (t)) \dot{\alpha}^a (t)
\left (  \hat {\bf T}[\beta \circ \alpha ] -
\hat {\bf T}[\beta \circ \alpha^{-1} ]
\right )
\ff
and
\f
\left [ \hat{B}^a [\beta ](s) , \hat T [ \alpha ]  \right ]
\equiv
\int dt \delta^3 (\beta (s), \alpha (t)) \dot{\alpha}^a (t)
\left (  \hat {B}[\beta \circ \alpha ] -
\hat {B}[\beta \circ \alpha^{-1} ]
\right )
\ff
The commutation relations not so far defined may be
worked out from these definitions and naturally
extend the usual $T$ algebra.

Finally, we may note that if we measure only observables
constructed from local fields there will be no way to tell
which womhole mouth a particular mouth is connected to.
This means that we should symmetrize or antisymmetrize
over the possible identifications among the points.  As dressed
wormholes with one loop passing through them behave, in the
neighborhood of each mouth, as spin $1/2$ excitations,
it is likely that the spin statistics theorem (which has
been generalized to the case of diffeomorphism invariant
theories\cite{spinstatistics}) applies to this case and requires that
antisymmetrization be chosen.  Thus, I will choose here this
option and restrict attention to
the antisymmetrized states,
\f
<\alpha ; [x_1, ..., x_{2n} ] | \equiv {1 \over {2n!}}
\sum_{{\cal P} [ i_1,...,i_{2n}   ]} (-1)^{\cal P}
<\alpha_{\cal P}  ;  (x_{i_1}, x_{i_2} ) ... | .
\ff
where ${\cal P}$ are all the permutations of the indices among the
first $2n$ integers.
We may note that whenever a loop in $\alpha$ is an open segment
in $\Sigma$ (that is then closed by passage through a mouth) the
routing of the new loop $\alpha_{\cal P}$  in each term
follows from the identifications in that permutation of the points.
We may note that when such states are dressed
(so there are no naked wormholes), they are defined already from
the open curves in $\alpha$, so they may be denoted simply
by $<\alpha |$ with the understanding that there are wormhole
mouths
associated with each open end in $\alpha$ and the identifications
among them are summed over antisymmetrically.
Finally, we may denote the
spaces of these antisymmetrized dressed states as
${\cal H}^\prime_{[x_1,...,x_{2n}]}$.

We may note that once we impose these antisymmetrizations
there may be at most two loops entering any wormhole mouth.
This is  most easily seen in the connection representation, where
a wormhole with $n$ lines coming from it may be seen an
object with $n$ independent spinor indices of the form,
$U_{\alpha_1 A_1}^{\ \ \ A^\prime_1}...
U_{\alpha_n A_n}^{\ \ \ A^\prime_n}$, where the $\alpha_i$
are loops exiting from the other ends of the (possibly
several wormholes that share a common mouth.  However,
by the antisymmetrization of wormholes, this must be summed
antisymmetrically over the $n$ loop segments, $\alpha_i$,
so that it vanishes when $n$ is greater than two.

Finally, it is particularly simple to write operators
that remove a
pair of wormhole  acting on these antisymmetrized states.
If
we break the generic multiloop $\alpha$ into
components $\beta_i$
that pass through wormholes, with the remaining, closed,
loops denoted
by $\gamma$, then we may define for each open segment,
$\alpha$ an operator $\hat X[\alpha ]$, by
\begin{eqnarray}
<\beta_1,...,\beta_n, \gamma | \hat{X}[\alpha ]
& \equiv &
\sum_i \sum_j <\beta_1,...,
\beta_i \circ \alpha \circ \beta_j,...,\gamma |
\delta^3 (\alpha (0), \beta_i (1))
\delta^3 (\alpha (1), \beta_j (0))
\nonumber \\
 && + 3 \ \ \mbox{other \ terms}   \nonumber  \\
&& +\sum_i <...\beta_{i-1},
\beta_{i+1},...,\gamma, \beta_i \circ
\circ \alpha |
\delta^3 (\alpha (0), \beta_i (1))
\delta^3 (\alpha (1), \beta_i (0))
\nonumber \\
&& + 1 \ \ \mbox{other \ term}
\end{eqnarray}
where $\beta_i \circ \circ
\alpha $ denotes the closed loop with $\alpha$ and $\beta$
joined on both ends, and in the other terms correspond to
switches of orientation of the loops.

\section{Connection with fermions at a kinematical level}

In the last section we introduced the kinematics of pure
quantum general relativity, without matter, but with
topology change, in the form of addition, motion of, or removal
of minimalist wormholes.   In this section I will show
that this theory is identical to the kinematics of quantum
general relativity coupled to one flavor of Weyl fermions.
This will be done by establishing an isomorphism between
the operator algebras of the
two theories.  This implies, and is equivalent to the demonstration
of an isomorphism between the state spaces of the two theories.

The addition of fermions to general relativity or Yang-Mills
fields in the loop representation has been described
several places\cite{loop-fermions,carlohugo}.
In this paper Weyl fermions
will be denoted $\psi^A (x)$ and their canonical
momenta (which are densities) by $\psi^\dagger_A (x)$
We assume the standard
anticommutation relations,
\f
\left [ \psi^A (x), \psi^\dagger_B (y) \right ]_+ =
\delta^{\ \ A}_B
\delta^3 (y,x)
\ff
The gauge invariant states may be constructed in the
connection representation by including fermions always
in the combinations
\f
W[\alpha ] = \psi^A (\alpha (0)) U_{\alpha A}^{\ \ B}
\psi_B (\alpha (1))
\ff
where $\alpha$ is any curve and $U_\alpha$
is the standard path ordered
holonomy operator in the spinor representation.
The connection representation may then be extended to
include states of the form
\f
\Phi [A,\psi ] = \Phi [ T[\gamma ], W[\alpha ] ]
\ff
where the dependence on the right hand side is assumed to be
analytic.  The corresponding states in the loop representation
are functions of closed and open loops, and are spanned
by the basis $<\alpha |$ where $<\alpha |$ is a multiloop
containing both closed and open loops.

The correspondence between these states and those described
in the previous section is obvious.  Both are spanned by
the basis $<\alpha |$, where closed and open lines are included
and no open line is repeated.  To show the isomorphism
explicitly is simplest at the level of the operator algebra, the
basic idea of the correspondence is that
\f
\psi^A (x) \psi^B (y) \Longleftrightarrow -
\epsilon^{AB} \hat A (x,y)
\ff
The antiymmetrization among wormholes guarantees that
the antiymmetrization among fermions is respected.  For
example, we have,
\begin{eqnarray}
\psi^A (x) \psi^B (y) \psi^C (z) \psi^D (w)
& \Longleftrightarrow  &
\epsilon^{AB} \hat A (x,y)\epsilon^{CD} \hat A (z,w)
-\epsilon^{AC} \hat A (x,z)\epsilon^{BD} \hat A (y,w)
\nonumber  \\
&& -\epsilon^{AD} \hat A (x,w)\epsilon^{BD} \hat A (z,y)
\end{eqnarray}
In terms of gauge invariant operators, we may write this
correspondence as,
\f
\hat{W}[\alpha ] \Longleftrightarrow  \hat{\bf T}[\alpha ]
\ff
The operator $\hat B[\alpha ]$ that moves wormholes
may then be put into correspondence with an operator
containing one $\psi^A$ and one $\psi^\dagger_A$,
\f
Y(\alpha ) \equiv \psi^A (\alpha (0)) U_{\alpha A}^{\ \ B}
\psi^\dagger_B (\alpha (1))
\Longleftrightarrow \hat{B}[\alpha ]
\ff

It is straightforward to show that these operators (on the
spinor side), together with the standard closed loop $T^0$
and $T^1$ operators have a closed
algebra\cite{loop-fermions,carlohugo}.  It is
similarly a straightforward exercise to compute the
algebra the $\hat{\bf T}[\alpha ]$'s, $\hat B [\alpha ]$'s
generate together with the standard closed loop
$T$'s on the state space defined in the previous section
and show that these two algebras are isomorphic.  For
brevity I will omit the details of these
standard calculations here.

Once the isomorphism is established one may further show
that the wormhole removal operator $X[\alpha ]$
corresponds to the action of the operator
$\psi^{A  \dagger} (\alpha (0)) U_{\alpha A}^{\ \ B}
\psi^\dagger_B (\alpha (1))$.

\section{Equivalence of fermions and wormholes at the
diffeomorphism invariant level}

Now that we have established the equivalence of the algebras
of wormhole and fermion creation operators we must show
that both sets of operators behave the same way under
diffeomorphisms.  As, for fermions, if $\hat U[\phi ]$ is the
operator that generates the diffeomorphism, we have of course,
\f
\hat{U}[\phi ] \psi^A (x) \psi^B (y) =
\psi^A (\phi \circ x) \psi^B (\phi \circ y) \hat{U}[\phi ]
\ff
But the same relation is satisfied by the operator
that creates a wormhole, because by the definition
of $A(x,y)$
\f
\hat{U}[\phi ] \hat A(x,y) =
\hat A(\phi \circ x, \phi \circ y) \hat{U}[\phi ]
\ff
Because the algebra, $C^\infty (\Sigma_{(x_1,y_1)...(x_n ,y_n)})$,
of
functions that
acts on each of the "almost manfolds"
$\Sigma_{(x_1,y_1)...(x_n ,y_n)}$ is a subalgebra
of $C^\infty (\Sigma) $, each element,
$\phi$, of the diffeomorphism
group $Diff (\Sigma )$ has a well defined action
in which $C^\infty (\Sigma_{(x_1,y_1)...(x_n ,y_n)})$
is mapped to
$C^\infty (\Sigma_{(\phi \circ  x_1, \phi \circ y_1)...
(\phi \circ x_n ,\phi \circ y_n)})$.  Similarly,
the operator $\hat U [\phi ]$ defines a one to one
map from
${\cal H}_{(x_1,y_1)...(x_n ,y_n)}$ to
${\cal H}_{(\phi \circ  x_1, \phi \circ y_1)...
(\phi \circ x_n ,\phi \circ y_n)}$ such that
$\hat{U}[\phi ]  \hat{P}[(x_1,x_2)...(x_{2n-1}x_{2n})]=
\hat{P}[(\phi \circ  x_1, \phi \circ y_1)...
(\phi \circ x_n ,\phi \circ y_n)]\hat{U}[\phi ]$.
Thus, the diffeomorphism
group of the original manifold $\Sigma$ acts on
the whole state space, ${\cal H}_{total}$, of the
theory.  This allows us to define the diffeomorphism
constraint on ${\cal H}_{total}$ as
$\hat{D}(v) \equiv -{d \over dt} \hat U [ \phi_t ]$, where
$\phi_t$ is the one paramater group of diffeomorphisms
generated by the vector field $v^a$.

We may then impose the
diffeomorphism constraints for the original
manifold on the quantum theory, and restrict
ourselves to states $\Psi$ that satisfy,
\f
\hat U [\phi ] \circ \Psi = \Psi
\ff
for all $\phi \in Diff (\Sigma )$.  Such states
live in a linear subspace of ${\cal H}_{total}$ we will call
${\cal H}_{diffeo}$.  It follows, from the action
of $\hat U[\phi ]$ and the antisymmetrization
condition, that each of these states is
defined by its values on one of the subspaces
${\cal H}_{(x_1,y_1)...(x_n ,y_n)}^\prime$ for each $n$.
We may then write,
\f
{\cal H}_{diffeo}^\prime = \sum_{n=0}^{\infty}
{\cal H}_{diffeo,n}^\prime
\ff
where ${\cal H}_{diffeo,n}^\prime$ contains those diffeomorphism
invariant states with $n$ wormholes.  This space then has
a basis which is given by all diffeomorphism equivalence
classes of closed curves on $\Sigma$ that thread $n$
minimalist wormholes.  These classes are equivalent to
all diffeomorphism equivalence classes of closed and open
loops on $\Sigma$, where there are $2n$ distinct end points.
But this is then exactly the same as
the diffeomorphism equivalence classes of the loop states
of the theory which includes fermions, so that the two theories
are isomorphic also at the level of the diffeomorphism
invariant states.

\section{Dynamical equivalence of fermions and wormholes}

When we add fermions to general
relativity, we must add terms to the hamiltonian and
diffeomorphism constraints of the form\cite{ART}
\f
{\cal C}^\psi (x) = ({\cal D}_a \psi )_A
s\tilde{E}^{a \ AB}\psi^\dagger_B
\ff
\f
D_a^\psi =({\cal D}_a \psi )^A \psi^\dagger_A
\ff
We could use the correspondence between fermionic and
wormhole operators to write the corresponding operators
in the wormhole language.  However, if we are really going to
implement the idea that the matter is not fundamental, but is
only a consequence of space having non-trivial topology, what
we would like is that the theory of {\it pure gravity} on
the theory
including wormholes be {\it dynamically}
equivalent to the theory
of gravity with matter added. There is indeed evidence that
this may be
exactly the case, as I will now show.

To calculate the action of the fermionic term of the
hamiltonian constraint we may regulate it using the
standard methods\cite{review-ls,review,weaves}.
Following work of Morales and Rovelli\cite{carlohugo},
we may write the regulated version of (25) as
 \f
\hat{\cal C}^\psi_\delta (N) = \int d^3x N(x)
\int d^3y \int d^3z
f^\delta (x,y) f^\delta (x,z)
({\cal D}_a \psi )^A(x) U_{\gamma_{x,y} A}^{\ \ \ B}
\tilde{E}^{a \ C}_{B}(y) U_{\gamma_{y,z} C}^{\ \ \ D}
\tilde{E}^{a \ AB}\psi^\dagger_D (z)
\ff
where
$ f^\delta (x,y) = {3 \sqrt{h} \over {4 \pi \delta^3 }}
\Theta ( \delta - | x-y|_h) $ is the
standard smearing function\cite{weaves}
($\Theta$ is the step function) defined with respect
to an arbitrary background flat metric $h_{ab}$ and
$\gamma_{x,y}$ is an arbitrary curve, defined with
reference to $h_{ab}$ from $x$ to $y$.

It is not difficult to show \cite{carlohugo}
that the action of this on
open loop states associated with pairs of fermions is, to
leading order in $1/\delta$ equivalent to the action of
a diffeomorphism that acts only at the ends of the open
loops to extend them outwards along their
tangent vectors.  That is,
if $\alpha$ is an open loop
\f
< \alpha | \hat{\cal C}^\psi_\delta (N) = <\alpha |
\int d^3 x \hat{V}^a_h \hat{D}_a^\psi
\ff
where $\hat{V}$ is an operator, defined in the loop
representation, and dependent on the background metric
such that
\f
<\alpha | \hat{V}^a_h = {1 \over \delta^2}<\alpha | v^a_\alpha
\ff
where $v^a_\alpha$ is a vector field, defined with
respect to the background metric $h_{ab}$, such that
\f
v(\alpha (1))^a = {3Nh(\alpha (1))  \over 4\pi}
{\dot{\alpha}^a(1) \over |\dot{\alpha} (1)|}
\ff
with the same equation (with a negative sign)
holding at the other
end point.

We may note that as $\hat{V}^a_h$ is an operator, this does not
imply that arbitrary diffeomorphism invariant states are in the
kernel of $\hat{\cal C}^\psi_\delta (N) $.  But what is
remarkable is that a similar equation holds in the case of the
pure gravity theory with wormholes.  This is because the
action
of the hamiltonian constraint on any loop that goes through the
wormhole is exactly like the action on a loop with a
nondifferentiable
point, and this action has been shown previously
to be proportional,
to leading order in the regulator, $\delta$, to the
action of a
diffeomorphism\cite{gambini,tedlee}.  In particular, up to a
regularization
dependent numerical factor, we have, with $\beta$ a loop
that goes through a wormhole identifying the points $x$ and $y$
\f
< \beta | \hat{\cal C}^{pure \ gravity}_\delta (N) = <\beta |
\int d^3 x \hat{W}^a_h \hat{D}_a^{pure \ gravity}
\ff
where now $\hat{W}$ is a background dependent
operator that acting at a nondifferentiable point produces
the same vector field as in the fermion case
\f
<\beta | \hat{W}^a_h = {1 \over \delta^2}<\beta | v^a_\beta .
\ff

Thus, we see that the pure gravity hamiltonian constraint on
the states of closed loops that transverse minimalist wormholes
has, at least to leading order in the regulators, precisely
the same action as the added fermionic term in the hamiltonian
constraints has on open lines that correspond, in the loop
representation, to a pair of fermions connected by an holonomy
element.   However, in each case, it can be shown that the
subleading terms can be eliminated so that the space
of solutions is determined only be the leading terms
in the action of the Hamiltonian constraint.  This then
establishes the dynamical equivalence of the
two theories for simple states of the form we have been considering.

It remains to show the equivalence for all states, which means
extending what has been done here to states in which fermions
or wormholes sit at intersection points of other loops.  Work
in this direction is in progress.

\section{Conclusions}

What we have found in this paper is that if the theory of the world
is generally relativity coupled to one Weyl fermion field, that
world may be
indistinguishable from a world whose degrees of freedom
are only the geometry and topology of spacetime, at least if the
possibilities for topology change are restricted to
minimalist wormholes, as done here.
The main thing that needs to be done is to extend this work
by showing that the dynamical equivalence holds for arbitrary
states involving fermions.  The kinds of states that need to
be considered are those in which the fermions sit at intersection
points of other loops.

Beyond this, what we would like then
to ask is whether this result can be extended to a more realistic
theory, in such a way that we might really be able to believe that
matter is a manifestation of the topology of space.

I will then close with several questions and comments about
extending this result to the context of a meaningful theory.

1)  Can the result be extended to incorporate Yang-Mills fields?
We may consider the theory of general relativity coupled to
Yang-Mills theory in the Ashtekar
formulation\cite{ART} with some compact gauge group $G$, taken
again in the case that the theory admits creation and annihilation
of minimalist wormholes, but contains no other matter.  Let us
quantize the theory in either the double connection representation,
in which the states are functions both of the gravitational
connection $A_{aAB}$ and the Yang-Mills connection
$a_{aIJ}$, or
in the corresponding loop representation.  Taking again the
same space of states, in which the connection representation
states are again products of traces of holonomies,
with the Yang-Mills holonomies taken in some representation
$\cal R$ of $G$.   It follows directly from what has been
shown here  that the wormholes
with one gravitational line and one gauge line will appear like a
multiplet of Weyl fields, in the representation $\cal R$ of the
Yang-Mills gauge group.   What has not been shown, however
is that the couplings of these fermions to the Yang-Mills fields
will be minimal, as there is a possibility of non-minimal couplings
arising from additional terms coming from the regulated Yang-Mills
terms in the Hamiltonian constraint.

2)  One can  also attempt to extend the present result to the
 non-trivial unifications of gravity and Yang-Mills theory that
Peldan and Chakraborty have
shown\cite{peldan} arise when the group in the
Ashtekar form of the constraints is taken to be larger than
$SU(2)$.  In the cases in which the larger symmetry is broken
to $SU(2) \times G$, in which Einstein-Yang-Mills theory emerges
in the low energy limit, wormholes will again behave like Weyl
fermions in some representation of $G$.  Again, there may be
non-minimal couplings at the Planck scale.

3)  The same remarks hold in the simpler case in which
Gambini and Pullin have shown that Einstein-Maxwell theory
emerges from the Ashtekar formalism by extending the internal
gauge group to $U(2)$\cite{unify}.

4) It is interesting to note that, in any of these approaches,
it will be impossible to generate any global symmetries
of the fermions that
are not at the shortest distances local symmetries.

5)  The basic correspondence shown here will work also for
quantum gravity in $2+1$ dimensions, in the case that the
pure gravity Hamiltonian constraint is taken  to be of the
form $FEE$, as in $3+1$.  On the other hand, Bruegmann and
Varadarajan have shown that there is a choice of representation
 space for $2+1$ quantum gravity in this form such that
the space of physical states is isomorphic to the state space
of the Witten form
(where the constraints are $F=0$)\cite{berndmadhavan}.
This result implies that $2+1$ gravity coupled to one Weyl
fermion field may be
an exactly solvable system whose state
space is exactly the subspace of the state space of $2+1$
gravity taken over all toplogies, in which the condition
of antisymmetrization over wormhole mouths has been
imposed.

6)  The correspondence shown here holds in the case of
states of the form of
\f
\Psi [A,\psi ] = e^{{1 \over G^2 \Lambda }
\int_\Sigma Y_{\mbox{Chern \  Simon}}(A)}  \Phi
\left [T[\gamma, A] , W[\alpha , A, \psi ] \right ]
\ff
which may be interpreted, in a semiclassical expansion,
as an expansion around the DeSitter vacuum\cite{chopinme}
with cosmological constant $\lambda$.

6)  We may also ask whether the correspondence between
fermions and topology can be extended to the more studied
case in which the wormholes are smooth manifolds.   It seems
that this is likely to be the case, at least in the semiclassical
limit, given that any kind of wormhole with one Wilson loop
emerging must look at low energies like a Weyl fermion.
In this case there are likely to be non-minimal couplings at
the scale of the wormhole size, so that the exact correspondence
found here is unlikely to be found for more smooth
wormholes.

Finally, one may ask whether the result found here is, in some sense,
trivial, given that minimalist wormhole mouths must behave
kinematically like Weyl spinors.  It would certainly not
be  surprising to find
that when one takes into account the relation between spin and
statistics  wormhole mouths behave {\it in the low energy
limit in which one expands around a state corresponding
to a classical geometry}  dynamically as Weyl spinors.
However, what is not required by this
consideration is that the correspondence be exact at the operator
level of the full non-perturbative theory, as there is no reason
coming from the semiclassical anaylsis or the spin statistics theorem
that
non-minimal Planck scale couplings might not
 appear when
wormholes were interepreted in terms of spinor fields.  The
fact that this is the case for general relativity, at least
for a large class of states gives us perhaps reason
to take seriously both the
interpretation of fermions as wormholes and
the conjecture that the fundamental nonperturbative
dynamics at the
Planck scale is the connection dynamics of general relativity.

\section*{ACKNOWELDGEMENTS}

I would like to thank Louis Crane and Carlo Rovelli for discussions
and encouragement about this idea, Hugo Morales for discussions
about fermions in the loop representation and the members of
the Center for Gravitational Physics and Geometry, especially
Don Marolf,  for spirited
criticism.  This work was supported by the National Science
Foundation under grants PHY 9396246 to The Pennsylvania
State University and Syracuse University.

\end{document}